\pdfoutput=1
\documentclass[11pt]{article}
\usepackage{acl}
\usepackage{times}
\usepackage{latexsym}
\usepackage{subcaption} 
\usepackage[T1]{fontenc}
\usepackage[utf8]{inputenc}
\usepackage{booktabs}
\usepackage{authblk}
\usepackage{microtype}
\usepackage{enumitem}
\usepackage{multirow}
\usepackage{inconsolata}
\usepackage{graphicx}
\expandafter\def\expandafter\UrlBreaks\expandafter{\UrlBreaks
  \do\a\do\b\do\c\do\d\do\e\do\f\do\g\do\h\do\i\do\j%
  \do\k\do\l\do\m\do\n\do\o\do\p\do\q\do\r\do\s\do\t%
  \do\u\do\v\do\w\do\x\do\y\do\z\do\A\do\B\do\C\do\D%
  \do\E\do\F\do\G\do\H\do\I\do\J\do\K\do\L\do\M\do\N%
  \do\O\do\P\do\Q\do\R\do\S\do\T\do\U\do\V\do\W\do\X%
  \do\Y\do\Z}

\title{NewsHomepages: Homepage Layouts Capture Information Prioritization Decisions}

\author[1*]{\bf Ben Welsh}
\author[2*]{\bf Arda Kaz}
\author[2]{\bf Michael Vu}
\author[2]{\bf Naitian Zhou}
\author[3*]{\bf Alexander Spangher}
\affil[1]{Thomson Reuters}
\affil[2]{University of California, Berkeley}
\affil[3]{University of Southern California}
\affil[ ]{\texttt{spangher@usc.edu}}

\begin{document}
\maketitle
\let\thefootnote\relax\footnote{* indicates co-first authorship}
\begin{abstract}

Information prioritization plays an important role in how humans perceive and understand the world. Homepage layouts serve as a tangible proxy for this prioritization. In this work, we present NewsHomepages, a large dataset of over 3,000 new website homepages (including local, national and topic-specific outlets) captured twice daily over a three-year period. We develop models to perform pairwise comparisons between news items to infer their relative significance. 
To illustrate that modeling organizational hierarchies has broader implications, we applied our models to rank-order a collection of local city council policies passed over a ten-year period in San Francisco, assessing their ``newsworthiness''. Our findings lay the groundwork for leveraging implicit organizational cues to deepen our understanding of information prioritization.

\end{abstract}

\section{Introduction}

The way humans prioritize and organize information is central to attention and understanding. Prior work has explored the need for effective prioritization to manage information overload \cite{miller1956magical} and including case-studies of prioritization approaches \cite{rosenfeld2002information, hays2018analysis}. However, larger scale analyses, specifically with an aim towards predictive modeling, are often limited by lack of available data and methods.

\begin{figure}[t]
    \centering
    \includegraphics[width=0.8\linewidth]{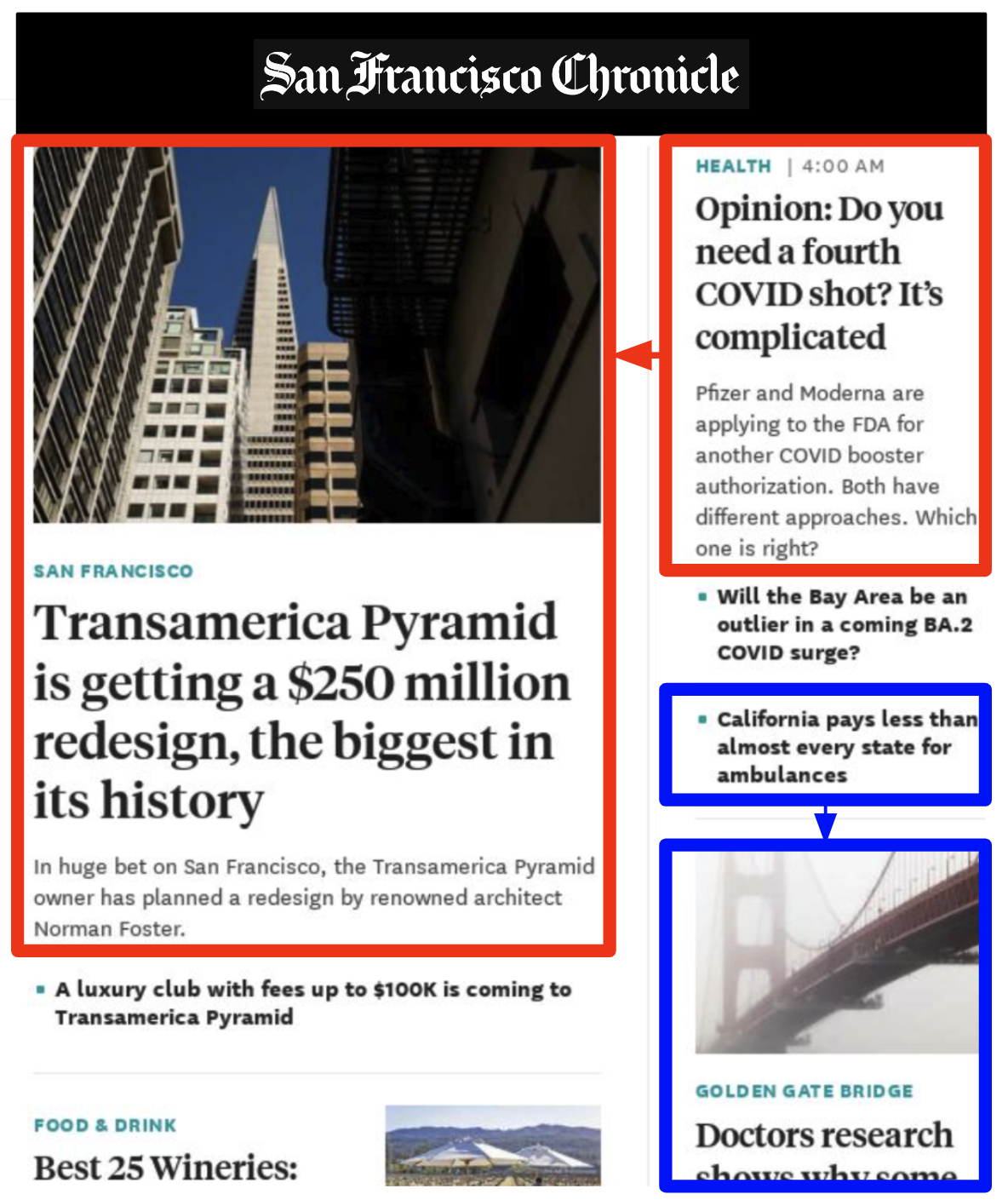}
    \caption{Two ``newsworthiness'' signals that editors make to guide reader attention are shown above. (1) \textcolor{red}{\textbf{Position}} (i.e. articles that are placed above, $\uparrow$, and left, $\leftarrow$ relative to other articles are more important \newcite{hays2018analysis}). (2) \textcolor{blue}{\textbf{Size}} (i.e. articles that are larger than other articles are more important) (3) \textcolor{blue}{\textbf{Graphics and Font}} (i.e. articles with graphics and images are more important). We release \textit{NewsHomepages}, a large dataset of over 3,000 homepages, collected twice-daily over three years, study information prioritization in this setting. We show can model these decisions at scale and demonstrate the usefulness of these models on two downstream tasks.}
    \label{fig:cover_photo}
\end{figure}

In this work, we seek to rectify this by focusing on a prominent source of publicly available, highly curated examples of prioritization decisions. News organizations' homepages are meticulously crafted by professional human editors: their layouts offer manifestations of the news organization's prioritization principles \cite{boukes2022newsworthiness}. We use homepage layouts to study information organization at scale. We present \textit{NewsHomepages}, to our knowledge the first large-scale homepage layout dataset, consisting of over 363k full page screenshots, HTML, extract links and other metadata from over 3,000 websites, which we captured twice daily over a three-year period, from local, magazine, and national news homepages. With this dataset, we ask: \textit{how well can we model the editorial decisions behind homepage layout? What do these decisions tell us about the priorities of the newspaper and the world?}

To answer these questions, we develop models to process and learn from our dataset at scale. First, we develop a novel bootstrapping approach to robustly parse homepage screen captures and determine precise spatial positions of news articles on homepages. Then, we train models to perform pairwise comparisons between articles based on their positions within layouts. By interpreting positional cues as indicators of significance, our models infer the relative importance of information.

The ability to learn these decisions with high accuracy indicates the robust signal this dataset captures. We conducted two downstream experiments utilizing these models; in each, we applied our pairwise comparison models to rank-order a list of text. In the first experiment, we use models trained on the newsworthy judgments of one news outlet to sort the news articles of another. We find surprising nuances: for instance, despite \textit{Breitbart} (a right-leaning outlet) being topically dissimilar to \textit{Mother Jones} (a left-leaning outlet), their newsworthy preferences are among the most correlated of outlets we studied. In the second analysis, we used each outlet's newsworthiness judgments to rank-order local city council policies passed in San Francisco \cite{spangher2023tracking}. By assessing these policies' ``newsworthiness'' through the lens of organizational prominence, we highlight how implicit cues in information structure can transfer between domains and serve as useful tools (in this case, to help human journalists find relevant story-topics). Our contributions are as follows:
\begin{itemize}
    \item We introduce NewsHomepages, large-scale dataset of hompage layouts from over 3,000 local, national and topic-specific news organizations, and develop a robust, weakly-supervised method to parse articles.
    \item We demonstrate that editorial principles behind information prioritization can be learned, by training pairwise comparison models to predict article size and positionality.
    \item We show via two case-studies -- (1) newsworthiness agreement between outlets, and (2) generating newsworthiness rankings for non-news corpora -- that such learned principles can generalize beyond the corpora we study and provide useful tools for end-users.
\end{itemize}

This work opens new avenues for exploring how implicit cues in digital environments reflect latent priorities and organizational principles. By modeling these cues, we can gain richer insights into the mechanisms that influence human perception.

\section{Background and Dataset}
\label{sct:dataset_collection}

Visual cues for editorial importance on homepages have a deep history in the design principles of physical newspapers~\cite{Barnhurst2001} and result from deliberate editorial processes. At \emph{The New York Times}, for example, top editors and designers convened daily in the renowned \emph{Page One} meeting \cite{usher2014making} to determine the most important articles for the print newspaper the next day\footnote{Terms like ``above the fold'' emerged to signal story-importance (i.e. the story is above the point at which the newspaper folds, so it is seen on newsstands)}. In the digital era, meetings like this evolved into \emph{Homepage Meetings} ~\cite{Sullivan2016}, influencing the design and content placement on the website's homepage for the upcoming day. As such, homepages continue to be distillations of professional judgement and priorities.

One visual cue editors use is \textbf{positional placement}, with articles positioned towards the top and left of a page considered more important~\cite{Nielsen2006}. This stems from observations that readers naturally begin scanning from the top-left corner \cite{bucher2006relevance}. Secondly, the \textbf{space} articles occupy is considered: larger articles or headlines are perceived as more important~\cite{Garcia1987}. In print media, prominence is conveyed through more column space; in digital media, longer headlines, featured images, and extended summaries. Finally, \textbf{graphics and design} also play a pivotal role in signaling the importance of news stories. Articles accompanied by photographs, videos, or other multimedia elements are often deemed more significant~\cite{Zillmann2001}. The use of capital letters, bold fonts, and color further enhances a story's prominence.

We find few large-scale computational analyses studying these attributes. To enable a more precise study of editorial judgement, we construct a large corpus of news homepage layouts, over which we can track these indicators of relative importance.


\subsection{Compilation of News Homepages}

We compiled a list of 3,489 news homepages, as of the time of this writing, which we scraped twice daily on an ongoing basis over a period of three years. From 2019-2024, we have collected a total of 363,340 total snapshots. Our dataset collection is actively maintained and facilitated by a large contributing community of over 35 activists, developers and journalists. We collect homepages from national news outlets (e.g., \emph{The New York Times}, \emph{The Wall Street Journal}), state-level news outlets (e.g., \emph{San Francisco Chronicle}, \emph{Miami Herald}), as well as local and subject-matter-specific news sources. Table \ref{tab:sample_homepages} provides a sample of the different categories of news homepages included in our dataset, and a full list can be found in the appendix. Additionally, we collect homepages from news websites of over 32 countries in 17 languages (please see Tables \ref{ref:languages} and \ref{tab:countries} for a more detailed breakdown). This is an ongoing and expanding effort: we encourage contributors to add their own news homepages of interest using for our suite of tools to scrape.\footnote{For more information on how to contribute, please see: \url{https://github.com/palewire/news-homepages}. For all code and data associated with this project, see \url{https://github.com/alex2awesome/homepage-newsworthiness-with-internet-archive}.
} We hope to further diversify the news sources in the dataset that we collect.

\begin{table*}[t]
    \centering
    \begin{tabular}{l l}
        \hline
        \textbf{Category} & \textbf{Example Outlets} \\
        \hline
        National & The New York Times, The Wall Street Journal, NPR, Bloomberg \\
        State-level & San Francisco Chronicle, Miami Herald, Chicago Tribune \\
        Local & Sturgis-Journal, The Daily Jeffersonian, LAist, The Desert Sun \\
        Subject-specific & The Weather Channel, Chessbase, ESPN \\
        International & India Today, Ukrinform, BBC, Prensa Grafica, Japan Times \\
        \hline
    \end{tabular}
    \caption{Sample of News Homepages by Category}
    \label{tab:sample_homepages}
\end{table*}

\subsection{Data Collection Pipeline}

Our dataset collection runs in a \texttt{chron} job twice a day, and uploads data to Internet Archive. For each snapshot, we store the following information:

\begin{enumerate}[noitemsep,nolistsep,wide, labelwidth=!, labelindent=2pt]
\item \textbf{All links on the page:} We store a flat-list of hyperlinks on every homepage and associated text.
\item \textbf{Full-page screenshots:} We store JPGs of each complete homepage as we render it.
\item \textbf{Complete HTML snapshots (subset of pages):} For a subset of homepages, we save a compressed version of the webpage, including all CSS files and images, using SingleFile\footnote{\url{https://github.com/gildas-lormeau/SingleFile}, incidentally the same software that Zotero uses. In initial experimentation, we observed that capturing complete, compressed HTML snapshots was far more robust than capturing assets}.
\end{enumerate}

In addition to our Internet Archive storage,\footnote{\url{https://archive.org/details/news-homepages}} we also synchronize with Wayback Machine to store these homepages, providing a secondary backup and ensuring long-term preservation.

\section{Dataset Processing}
\label{sct:dataset_processing}

In order to robustly extract visual attributes for each article on a homepage (i.e. size, position, presence of graphics), we need to determine bounding boxes for all articles on a homepage. Examples of bounding boxes are shown in Figure \ref{fig:cover_photo}: each bounding box, also referred to as \textit{article card}, covers all information directly associated with that article.



Layout parsing is a well-researched field \cite{shen2021layoutparser, Li2020DocBank}. However, homepages present unique challenges due to their diverse structures: text of varying size, fonts, colors and images are easily perceived by humans. However, because none of the largest supervised datasets \cite{zhong2019publaynet} are specific to our task, we find that existing resources fail for parsing homepages. So, we bootstrap a supervised detection task.

\subsection{Bootstrapping a Bounding Box Detector}

Following other bootstrapping approaches \cite{amini2022self}, we: (1) develop a simple deterministic algorithm to generate candidate data, (2) apply a filtering step to exclude low-quality data, (3) use our high-precision dataset to train a more robust classifier. Figure \ref{fig:training-pipeline}, in the Appendix, provides an overview of the pipeline.

\paragraph{Step 1: Find Bounding Boxes Deterministically}

We design a deterministic algorithm, called the DOM-Tree algorithm, to start our bootstrapping process. At a high level, the algorithm traces each \texttt{<a>} tag in the Document Object Model (DOM) and extracts the largest subtree in the DOM that contains \textit{only a single \texttt{<a>} tag} (illustrated in Figure~\ref{fig:combined_algorithm_figures}, Appendix). This method can extract the maximal bounding box for each article, however it faces robustness challenges, for example, if a link exists \textit{within} an article card (e.g. a link to an authors page, as shown in Figure~\ref{fig:bounding_box_extraction}, Appendix.) 

We apply this algorithm to a subset of the NewsHomepages dataset, combining 15 homepages each from all outlets for which we have HTML files, JPEG snapshots, and hyperlink json files (approximately 15,000 homepages). Since each outlet typically maintains a consistent layout on their homepages across samples, we include more outlets for generalizability. 


\begin{table*}[t]
    \small
    \centering
    \label{tab:error_analysis_condensed}
    \begin{tabular}{llcccccc}
        \hline
        & & \textbf{FP\#1} & \textbf{FP \#2} & \textbf{FN \#1} & \textbf{FN \#2} & Total Errors  & \% Correct \\
        \hline
        \multirow{2}{*}{Challenge dataset} & DOM-Tree algorithm & 117 & 137 & 127 & 265 & 646 &  61.3\% \\
        & Detectron2 Model & 25 & 23 & 27 & 87 & 162 & 90.3\% \\
        \hline 
        \multirow{2}{*}{Clean dataset} & DOM-Tree algorithm & 12 & 20 & 0 & 13 & 45 & 97.1\%\\
        & Detectron2 Model & 15 & 24 & 0 & 18 & 57 & 96.3\% \\
        \bottomrule
    \end{tabular}
    \caption{Error analysis of bounding box detection methods comparing the DOM-Tree algorithm and a Detectron2 model across two datasets: the challenge dataset and the clean dataset. The challenge dataset is formed by selecting the bottom 10\% of articles based on the match between OCR-extracted text and retrieved link text (described in Section \ref{sct:dataset_processing} Step 2), while the clean dataset contains well-matched articles. Error types are divided into false positives (FP \#1: multiple articles in one box, FP \#2: no articles in a box) and false negatives (FN \#1: partially captured articles, FN \#2: articles not captured). As can be seen, the our trained model performs at par on the DOM-Tree algorithm in the clean settings and is far more robust in noisy settings.}
    \label{table:bb_errors}
\end{table*}

\paragraph{Step 2: Filter Low-Quality Bounding Box Extractions}
We take several filtering steps to prevent ``drift'' \cite{amini2022self}. (1) First, we exclude non-news article links (e.g. log-in pages) by training simple text classifier to distinguish between URLs to news articles and others. We manually labeled over 2,000 URLs. The model achieves an accuracy of 96\%. (2) Then, we exclude bounding boxes that did not overlap highly with link text. We determine this by first rendering the HTML pages as images and overlaying bounding boxes, then running OCR to extract the bounding-box text. (3) Finally, we exclude bounding boxes with improperly rendered images\footnote{Likely due to errors in HTML extraction or dead links}. To address this, we again rendered HTML pages as an image and employed the YOLO object detection model \cite{yolov3} to compare these images to the JPEGs in our archive. If a screenshot was not within 80\% of the detection count of the archived snapshot, we discarded the snapshot. Overall, this multi-stage filtering process significantly reduced the number of boxes that did not correspond to actual articles and removed many websites that contained broken or corrupt data, enhancing the quality of the training data.

\paragraph{Step 3: Train a More Robust Classifier}

Now with our dataset in hand, we trained a Detectron2 model \cite{wu2019detectron2}. 
Our model uses ResNet-101 as a backbone with a Feature Pyramid Network (FPN) for extracting multi-scale features and Smooth L1 loss for bounding box regression. During training, we used a base learning rate of 0.02 with a linear warmup over the first 1000 steps.
We trained the model for 10,000 steps in total, with learning rate reductions after 5000 steps. A weight decay of 0.0001 and momentum of 0.9 were also employed. The training ran on 4$
\times$A40 GPUs for 24 hours. 

\subsection{Evaluation and Results}
\label{sct:eval}

To evaluate the quality of our bounding box detection, we conducted manual validation for four types of errors: 1) bounding boxes that contain multiple articles, 2) bounding boxes that contain no articles, 3) bounding boxes missing parts of an article, and 4) articles that are note captured. 



We used the OCR text-matching method, as described in Section \ref{sct:dataset_processing}, to identify particularly challenging homepages.\footnote{Figure \ref{fig:similarity_histogram}, in the Appendix, demonstrates the resulting histogram for the distribution of OCR-match scores across our dataset.}  We compared errors on the cleanest 10\% of homepages (Clean) and the least-clean 10\% (Challenge). As shown in Table \ref{table:bb_errors}, our computer vision model (Detectron2) significantly improved the accuracy of bounding box detection in contexts where the DOM-Tree algorithm struggles (the Detectron2 model had an Card Correct \% score of 90.3\% while the DOM-Tree had a score of 61.3\%). For Clean pages, the Detectron2 model performed similarly to the DOM-Tree algorithm, with error differences being minimal and both models achieving high accuracy (above 96\%). The combination of deterministic algorithms and machine learning techniques allow us to achieve a more robust extraction of article attributes from diverse homepage layouts. 

\begin{table*}[h!]
\small
\centering
\begin{tabular}{l ccc ccc}
\hline
 & \multicolumn{3}{c}{Size} & \multicolumn{3}{c}{Position x Size} \\
\cline{2-4} \cline{5-7}
\textbf{Model Name} & \textbf{F1} & \textbf{Precision} & \textbf{Recall} & \textbf{F1} & \textbf{Precision} & \textbf{Recall} \\
\hline
Flan-t5-base & 91.9 & 91.8 & 92.0 & 70.7 & 80.5 & 63.1 \\
Flan-t5-Large & 66.6 & 51.1 & 95.4 & 54.9 & 43.4 & 74.7 \\
Roberta Base & 91.0 & 91.7 & 90.4 & 64.9 & 78.1 & 55.6 \\
Roberta Large & 85.4 & 85.9 & 84.9 & 47.2 & 74.4 & 34.6 \\
Distilbert-Base-Uncased & 93.1 & 93.3 & 92.9 & 75.2 & 81.8 & 69.6 \\
\hline
\end{tabular}
\caption{Performance metrics on NYTimes data for different models}
\label{table:performance_metrics}
\end{table*}

\begin{table}[]
\small
    \centering
    \begin{tabular}{lrrrr}
    \toprule
    Outlet & Accuracy & F1 & Recall & Prec. \\
    \midrule
    phoenixluc & 57.1 & 70.3 & 57.4 & 90.7 \\
    newsobserver & 75.0 & 72.5 & 74.3 & 70.7 \\
    slate & 72.4 & 61.6 & 66.2 & 57.7 \\
    jaxdotcom & 75.2 & 63.4 & 65.5 & 61.4 \\
    arstechnica & 64.7 & 17.5 & 41.4 & 11.1 \\
    airwaysmagazine & 72.5 & 73.7 & 78.9 & 69.1 \\
    denverpost & 73.7 & 67.8 & 70.5 & 65.4 \\
    thedailyclimate & 82.0 & 80.9 & 81.3 & 80.6 \\
    breitbartnews & 68.9 & 22.8 & 54.7 & 14.4 \\
    foxnews & 67.3 & 38.6 & 55.6 & 29.5 \\
    motherjones & 71.4 & 63.0 & 68.7 & 58.2 \\
    thehill & 68.8 & 55.5 & 59.8 & 51.7 \\
    wsj & 70.0 & 48.0 & 52.0 & 44.6 \\
    \bottomrule
    \end{tabular}
    \caption{Performance metrics on a sampling of outlets, including on ones we used for the downstream experiments in Section \ref{sec:downstream_tasks}. }
    \label{tab:performance_sample}
\end{table}

\section{Homepage Modeling}
\label{sct:homepage_modeling}

Given precise layout information for the 363k homepages in our dataset, we arrive again at the core question of this research: how regular and predictable is an outlet's layout decisions? We initially hypothesized that modeling homepage placement would be challenging. As shown in Figures \ref{fig:annotated-homepage}, \ref{fig:homepage_visualization}, \ref{fig:homepage_tracking}\footnote{In Appendix \ref{app:news_homepage_layouts}} and described by \cite{angele2020metrics}, certain areas of many homepages lack clear editorial consistency. This introduces noise and makes it difficult to learn a uniform policy guiding article placement decisions. Further, learning a single set of policies is challenged by the changing news cycle; some days have lots of news while others have less.

\subsection{Modeling Approach}

A homepage is intended to present a collection of articles as a cohesive bundle; individual articles do not exist in isolation  \cite{tufte1990envisioning}. Predicting the placement of a single article without considering the context of other articles would be overly noisy and potentially ineffective \cite{salganik2006experimental}. Conversely, attempting to predict the placement of all articles simultaneously poses a combinatorial challenge that is computationally infeasible.

To address this issue, we formulate our modeling task as a pairwise preference problem. Specifically, we consider pairs of articles $(a_1, a_2)$ and train models to predict a binary preference variable $p$, where $p_o(a_1 > a_2) = 1$ if article $a_1$ is preferred over article $a_2$ for outlet $o$, and $p_o(a_1 > a_2) = 0$ otherwise.

We explore three variations of preference criteria for the preference variable, $p$:

\begin{enumerate} 
    \item \textbf{Size-based Preference}: We define $p_o(a_1 > a_2) = 1$ if article $a_1$ occupies more surface area on the homepage than article $a_2$, assuming that prominent articles are given more space \cite{lambert2005layout}. 
    \item \textbf{Position-based Preference}: We set $p_o(a_1 > a_2) = 1$ if article $a_1$ is placed in a more favorable location on the homepage than article $a_2$, such as higher up or more to the left, based on common reading patterns \cite{nielsen2006eyetracking}. 
    \item \textbf{Combined Size and Position Preference}: Here, $p_o(a_1 > a_2) = 1$ if article $a_1$ either occupies more surface area or is in a more favorable position than article $a_2$, particularly focusing on articles that are in the top 10
\end{enumerate}

To model these preference variables, $p$, we train a simple Transformer-based binary classifier, \texttt{distilbert-base(X)}, which classifies a text sequence $X$. Our model concatenates the input articles: \texttt{X=a}$_1$\texttt{<sep>a}$_2$ as input; the model learns to recognize the \texttt{<sep>} token as a boundary between the first and the second articles. 

\subsection{Modeling Variations}

We explored different modeling variations on the \textit{New York Times} homepages, as they have a variety of content, shares and functionalities on their site \cite{spangher2015building}. We test 5 different models: \{distilbert-base-uncased, flan-t5-base, flan-t5-large, roberta-base, roberta-large\} and constructed a training dataset of 74,857 article-pairs and a test dataset consisting of 18,715 datapoints consisting of pairs of NYTimes articles from same homepages.

We experienced exploding gradients in the flan-t5-large and RoBERTa-large models, motivating us to use a learning rate limit of 5e-5 for all the models, for the sake of equal comparison. We applied Parameter-Efficient-Fine-Tuning \cite{peft} on flan-t5-base, flan-t5-large, roberta-base, roberta-large models to minimize overfitting, as we had limited of datapoints.

The distilbert-base-uncased model outperforms other models (Table~\ref{table:performance_metrics}). We trained on 4xA40 GPUs and 16xA100 GPUs, and implemented gradient clipping after observing gradient explosion. 

\subsection{Dataset Selection and Processing}

From our list of 3,000 outlets, we select 31 outlets for detailed analysis. We selected well-known outlets in various categories, including different political leanings (left-leaning vs.\ right-leaning\footnote{As classified by MediaBiasFactCheck.com}), local and national levels, and varied subject matters such as science, chess and aviation. For each outlet, we collected between 200 and 300 homepage snapshots, resulting in 1,000 to 50,000 pairs of articles. We created an 80/20 train/test split and trained distilbert-base-uncased models for each outlet. We trained each model with 5e-5 learning rate limit, 3 epochs, 0.01 weight decay.

Each article in our dataset includes its textual representation as it appeared on the homepage. To enhance the reliability of our models, we undertake several data processing steps informed by preliminary experiments: (1) we only sample pairs of articles that are adjacent on the homepage, to curate preference pairs that are more likely to be challenging and topically similar. Secondly, we clean the textual representations by stripping out any times, dates, and formatting elements. We also remove author names to prevent the models from learning biases based on authors who might be favored by the organization. Please refer to Appendix \ref{app:dataset_details} for a detailed list of the outlets used and the specific number of data points associated with each.

\subsection{Results}

We show our results in Table \ref{tab:performance_sample}. While some models (e.g. \texttt{Breitbart}) perform noticeably poorly, we note that the majority of our models score above $f_1 > .6$. We do not find a significant correlation between model performance and training set size. We were surprised to observe the tractability of this task; this indicates that many of the concerns we had about noise were either handled by our preprocessing steps, or not as important as we believed. 

\section{Demonstrations}
\label{sec:downstream_tasks}

To evaluate the practical utility of our models, we design two downstream tasks: (1) analyzing newsworthiness agreement between publishers, and (2) using newsworthiness models to rank corpora of interest to journalists.

\subsection{Task 1: Newsworthiness Agreement Between Publishers}

In this task, we aim to rank-order lists of news items drawn from a larger pool of articles to calculate the agreement rates for newsworthiness decisions between different news outlets. Previous research has observed surprising overlaps in sentiment and preferences between right-leaning and left-leaning outlets \cite{gentzkow2010drives}, and we wish to quantitatively test this phenomenon using our preference models.

We selected 9 of the 31 outlets for which we trained preference models in the previous section. From each outlet, we sampled 1,000 articles, matching on variables such as topic, length, publication date, and other potential confounders. These 9 outlets were chosen because they represent a range of political viewpoints.

For each model $n_{o_i}$ (corresponding to outlet $o_i$), we used it to sort lists of 1,000 articles $\{a_1, a_2, \dots, a_{1000}\}_{j=1}^9$ from outlets $\{o\}_{j=1}^9$. In other words, the output of applying model $n_{o_i}$ to the article list from outlet $o_j$ is a fully sorted list $n_{o_i}(A_j)$. We used the size $\times$ position model for this experiment, as performance was similar to the size-only model, and we believed that the multivariable models capture more newsworthiness information than the single-variable models.

\begin{figure}[t]
    \centering
    \begin{subfigure}[]{\linewidth}
        \centering
        \includegraphics[width=\linewidth]{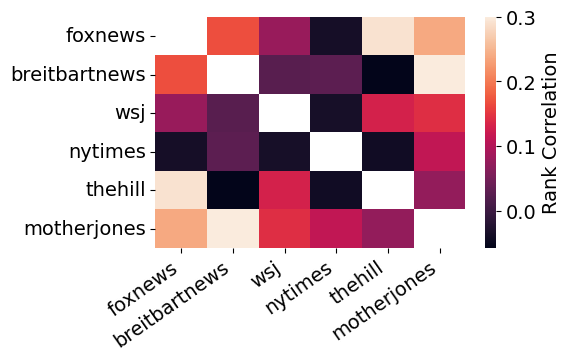}
        \caption{Kendall's $\tau$ correlation between the newsworthiness preferences expressed by preference models trained on different news outlets.}
        \label{fig:correlation_matrix}
    \end{subfigure}
    \vspace{0.5cm}
    \begin{subfigure}[]{\linewidth}
        \centering
        \includegraphics[width=\linewidth]{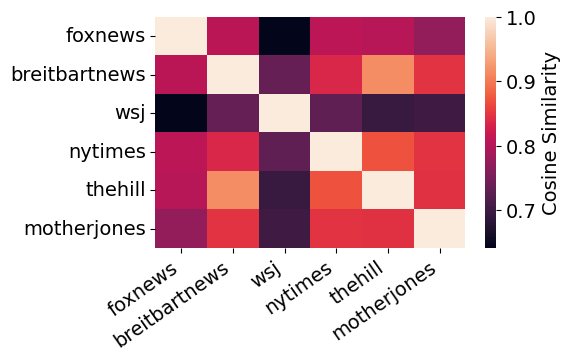}
        \caption{Cosine distance of average SBERT similarity between articles sampled from each outlet.}
        \label{fig:embedding_similarity}
    \end{subfigure}
    \caption{Comparison of Kendall's $\tau$ rank correlation (on newsworthiness judgements) and SBERT cosine similarity (on articles) across news outlets.}
    \label{fig:combined_similarity}
\end{figure}

We calculated Kendall's $\tau$, a correlation measure for ordinal data, between each pair of sorted lists $(n_{o_i}(A_k), n_{o_j}(A_k))$ for all $i, j, k$, and averaged the correlations across $j$. The resulting correlation matrix is displayed in Figure~\ref{fig:correlation_matrix}. Some surprising insights emerge from this analysis. Notably, \textit{Breitbart}, a right-leaning outlet, and \textit{Mother Jones}, a left-leaning outlet, have one of the highest rates of agreement.

To establish a baseline and ensure we are not merely capturing topic overlap (despite matching on topics), we conducted a simple SBERT embedding experiment \cite{reimers2019sentence}. We sampled a set of 100 articles per outlet, generated embeddings using SBERT, and averaged these embeddings to create single outlet-level embeddings, as shown in Figure~\ref{fig:embedding_similarity}. These embedding-level similarities align more closely with topical overlaps, indicating distinct right-wing and left-wing clusters with some overlap in between. 

Taken together, these results suggest that newsworthiness preference is a compelling and orthogonal variable for study beyond topical similarity.

\begin{table*}[ht]
    \small
    \centering
    \begin{tabular}{p{2cm}p{5cm}p{7cm}}
        \hline
        \textbf{Outlet} & \textbf{Top Policies LLM Summaries} & \textbf{Examples of Policies} \\ \hline
        Weather Channel & Environmental Policies, Public Health and Emergency Response, Infrastructure and Development & Reducing nutrient pollution from wastewater; Accepting grants for forensic science improvements \\ \hline
        Daily Climate & Environmental and Energy Policies, Urban Planning and Development & Agreement with North Star Solar; Building code enforcement \\ \hline
        Fox News & Community and Public Safety Policy, Education and Social Policy, Fiscal and Economic Policy & Appointment of individuals to advisory committees; Appropriating funds for San Francisco Unified School District; Developing materials on domestic violence \\ \hline
        Mother Jones & Social Policies, Environmental and Health Policies & Sanctuary City Protection; Urging Pardons; Edible Food Recovery and Organic Waste Collection \\ \hline
        Ars Technica & Infrastructure Policies & System Impact Mitigation Agreement; 6th St. Substation \\ \hline
        NYTimes & Social \& Cultural Awareness Policies, Labor \& Employment, Economic, Housing policies & Commemorative and Awareness Events; Labor Dispute Hearings; Affordable Housing Loans \\ \hline
        WSJ & Economic and Infrastructure Policies, Governance and Legislative Policies & Contract modifications; Bond sales; Ground lease agreements; Charter amendments concerning commissions and departments related to aging and adult services \\ \hline
    \end{tabular}
    \caption{Summaries of the top 10 most newsworthy policies published by the San Francisco Board of Supervisors, as ranked by models trained on 7 different homepages.}
    \label{tab:outlet_policy_case_studies}
\end{table*}

\subsection{Task 2: Surfacing Potentially Newsworthy Leads}

In this task, we explore how well these newsworthiness judgments transfer outside of the news domain. In this task, we build on the work of \citet{spangher2023tracking}. The authors introduced the task of \textit{newsworthiness prediction} as a detection and alerting system for journalists: utilizing a list of San Francisco Board of Supervisors' policies (a typical source of stories for journalists), they attempted to detect which policies were \textit{more newsworthy} in order to alert journalists. 

Here, we suspect that editorial cues from different homepages will help us surface especially newsworthy content based on the preferences of each outlet. We applied the models from each outlet to sort the list of Board of Supervisors' policies. Then, we selected the top 10 items from the ordered lists $n_{o_i}$ and used a large language model (LLM) to summarize the key points raised in each policy.\footnote{We used GPT-4 for this experiment.}

The LLM's summarization results and examples are shown in Table~\ref{tab:outlet_policy_case_studies}. Themes emerge, with subject-specific outlets like \textit{The Weather Channel} highlighting policies related to environmental issues. We presented these results to a group of journalists, and 81\% of respondents indicated they were impressed and would consider using such a system in their workflow.
These findings demonstrate the potential of our models to assist journalists in identifying newsworthy leads from large corpora of documents, thereby supporting investigative journalism and timely reporting.

\section{Discussion}

Our demonstrations show two core findings: first, editorial priorities and decision-making can be inferred simply by examining the layout decisions made on homepages. This decision-making is distinct from simple topic preferences, as we show. In fact, commonalities about decision-making can be observed between outlets that appear distinct topically. Second, newsworthiness judgements have potential to be used in tools for reporting. 

Stepping back, these results indicate that homepage editorial cues provide an interesting, novel angle for news analysis, as well as a tantalizing direction in newsworthiness detection \cite{spangher2023tracking, diakopoulos2010diamonds}. Both of these applications are premised by the assumption that editorial cues learned from one outlet's homepage can be transferred to other domains, be it \textit{another} outlet's articles, or non-news content. This is an important assumption: the intuitive findings that we have made in our demonstrations provide some degree of proof that this transfer is robust. 

We experimented with different ways of making this transfer even more robust. We attempted to train \textit{additional} models to serve as in-domain and out-of-domain classifiers, and then multiplied the probabilities: $\hat{P}_o(a_1 > a_2) = p_{o}(in\_domain | a_1, a_2) \times p_{o}(a_1 > a_2)$, where for outlet $o$, $p_{o}(in\_domain | a_1, a_2) = 1$ if $a_1$ belongs to $o$ while $a_2$ does not. However, our results in this direction were not more interpretable than the results we reported. Ultimately, without any ``gold truth'' about how an editor from outlet $o$ would rank an arbitrary list of strings, we will not have a conclusive measurement about our ability to replicate these measurements. In order to fully validate our observations, this appears necessary. Our results have to be taken with some important further caveats. While some newsworthiness models were impressively well-performing, many were not. Further exploration is needed to determine the causes. Additionally, despite the presence of non-English homepages in our dataset, we only tested with U.S.-based websites. We look forward to continuing to expand this work to address these concerns.

With these caveats in mind, we hesitate to draw firm conclusions. However, we feel the results of our modeling are encouraging enough to continue to address these concerns and run further manual experiments with editors. We imagine a future where editorial preferences made by professional editors can be used to routinely study importance and organization of content. We imagine insights being applied more broadly to build tools for journalists, improve webpage layouts that are currently automated, and even understand more fundamental components of the human psyche. 

\section{Related Work}

Understanding how information is prioritized and presented in news media has long been a subject of scholarly interest. This section situates our work within the broader literature on information prioritization, visual cues in editorial decision-making, layout parsing, modeling editorial judgments, and data-driven studies of news content and bias.

\paragraph{Information Prioritization and Newsworthiness}

The concept of \emph{newsworthiness} is central to journalism studies and media sociology. Classic works by \citet{galtung1965structure} introduced a set of news values that determine the selection and presentation of news stories. These news values have been revisited and updated by scholars such as \citet{harcup2001news} and \citet{harcup2016news}, who identified factors like relevance, timeliness, and unexpectedness as key determinants of newsworthiness. Prior research has explored the cognitive and organizational processes behind news selection. \citet{shoemaker1991gatekeeping} examined the gatekeeping role of editors and journalists in filtering news content. \citet{herman2021manufacturing} discussed how media organizations' structures influence news production and prioritization.

Our work contributes to this literature by providing a computational approach to inferring newsworthiness judgments from homepage layouts, offering a large-scale empirical perspective on editorial prioritization decisions.

\paragraph{Visual Cues and Editorial Decision-Making}

Visual presentation plays a crucial role in shaping readers' perceptions of news importance. Studies have shown that elements such as headline size, article position, and the use of images significantly affect reader attention and recall \cite{brooks2022art, nass1990study}.

Eye-tracking research has provided insights into how users interact with news websites. \citet{nielsen2006eyetracking} found that users' viewing patterns are influenced by the layout and design of webpages, with a tendency to focus on content positioned at the top and left areas of the screen. Similarly, \citet{bucher2006relevance} demonstrated that visual cues guide readers' attention and are integral to the perceived relevance of news stories. Our study builds on these findings by quantitatively modeling how spatial attributes of articles on homepages reflect editorial judgments about their significance.

\paragraph{Modeling Editorial Decisions}

Computational modeling of editorial decisions and news selection processes has been explored in prior research. \citet{arya2016news} proposed methods to learn news selection patterns from data, aiming to predict which news stories editors might choose to publish. Additionally, \citet{diakopoulos2010diamonds} investigated computational tools to assist journalists in identifying newsworthy data points within large datasets, emphasizing the role of algorithms in supporting editorial judgment.

Our approach differs by focusing on inferring editorial prioritization from the spatial arrangement of articles on homepages, rather than predicting content selection from textual features alone. By modeling pairwise preferences based on layout attributes, we provide a novel perspective on editorial decision-making.

\paragraph{Data-Driven Studies of News Content and Bias}

Large-scale analyses of news content have provided insights into media bias, framing, and agenda-setting. \citet{gentzkow2010drives} examined the factors influencing media slant in U.S. daily newspapers, using textual analysis to measure ideological positioning. Multiple research projects have aggregating and analyzing vast amounts of news content, enabling researchers to explore patterns in news coverage and discourse \cite{roberts2021media, misra2022news, silcock2024newswire, leetaru2013gdelt, spangher-etal-2022-newsedits}.

Our dataset, NewsHomepages, contributes to this line of research by offering a rich source of data on how news organizations present and prioritize information on their homepages, complementing textual analyses with spatial and visual dimensions.

\paragraph{Applications in Journalism and News Recommendation}

The intersection of computational models and journalism has led to the development of tools for news recommendation and content personalization. \citet{talebifard2014context} explored context-aware news recommendation systems that adapt to users' interests and reading behaviors. Moreover, computational approaches have been proposed to assist journalists in investigative work. \citet{spangher2023tracking, spangher-etal-2023-identifying, spangher-etal-2022-sequentially} developed methods to detect newsworthy events in local government proceedings, identify sources, and structure news articles.

Our work extends these applications by demonstrating that models trained on editorial cues from news homepages can be applied to rank a variety of corpora, possibly aiding in the design of future recommendation systems.

\paragraph{Visual Salience and Information Design}

The principles of information design and visual salience are relevant to our study. \citet{tufte1990envisioning} emphasized the impact of visual arrangement in effectively conveying information. In the context of web design, \citet{jiang2016determinants} discussed how layout aesthetics affect user engagement and perception.

Our findings align with these principles, highlighting how the spatial arrangement of content on news homepages serves as an implicit signal of importance, guiding reader attention and shaping information consumption.

\section{Conclusion}

In this work, we introduced NewsHomepages, a comprehensive dataset capturing editorial prioritization decisions through the lens of homepage layouts across thousands of news organizations. By applying weakly-supervised models, we demonstrated the ability to robustly infer editorial judgments through pairwise comparisons of article positioning.

\section{Contributions}

\textbf{Co-first author contributions}: Ben Welsh founded the \texttt{News-Homepages Internet Archive Project}, built and maintained the pipelines to perform data collection over a period of 8 years. Alexander Spangher was the primary writer, he was the advisor for Arda and Michael, and he performed the analysis in Section \ref{sec:downstream_tasks}. Arda Kaz did the modeling in Section \ref{sct:homepage_modeling}. \textbf{Secondary author contributions:} Michael Vu did most of Section \ref{sct:dataset_processing}. Naitian performed initial exploration in Section \ref{sec:downstream_tasks} and editing.

\section{Limitations}

This work, while advancing the study of editorial prioritization on homepages, comes with several limitations. First, the dataset, although large and diverse, predominantly focuses on English-language news outlets from the U.S., which may limit the generalizability of our models to international or non-English outlets. Despite the inclusion of some non-U.S. and non-English homepages, the models have not been explicitly evaluated on a broader range of languages or cultural contexts. This focus may overlook regional editorial conventions and biases that differ significantly from the U.S. context.

Another limitation is that the study focuses primarily on visual cues of newsworthiness, such as size, position, and graphical elements. While these cues are significant, they are not the only factors that influence editorial decisions. The models do not account for less visible but equally critical considerations, such as journalistic ethics, editorial mandates, or audience engagement metrics, which may influence homepage layouts but remain unquantified in this dataset.

Additionally, the weakly-supervised learning methods employed for layout parsing may struggle with more complex or irregular homepage designs. As described, the model had trouble generalizing to homepages with obscure HTML structures, leading to imperfect bounding box detections in some cases. This could result in misinterpretations of editorial significance, especially for websites with non-traditional or highly dynamic layouts.

Lastly, the results may have some bias due to the reliance on pairwise article comparisons. This method, while efficient, reduces the complexity of editorial decision-making into binary relationships, potentially overlooking more nuanced or multifaceted prioritization strategies that editors use in practice.

\subsection{Computational Budget}
The computational resources required for this project were substantial. The model training involved 4$\times$A40 GPUs for initial phases and 16$\times$A100 GPUs for more extensive model fine-tuning and deployment. Training the custom Detectron2 model alone took 24 hours, with additional time required for fine-tuning and model testing across multiple outlets. While the experiments could be completed within this budget, more extensive experiments across all 3,000 outlets would require significantly more resources, especially when scaling to different languages and regional variations.

\subsection{Use of Annotators}
The dataset used in this work was primarily compiled automatically through web scraping, with minimal manual annotation. However, annotators were employed to manually label URLs to distinguish between news articles and non-news articles during the preprocessing phase. A set of 2,000 URLs was manually labeled, contributing to a more refined and accurate dataset. The authors performed this task themselves. However, beyond this, no human annotators were used to manually assess newsworthiness, as the models relied on position and size as proxies for editorial decisions.

\bibliography{custom}

\clearpage
\appendix

\section{Dataset Details}
\label{app:dataset_details}

In this section, we present more detailed dataset statistics. In Table \ref{ref:languages}, we show the different languages collected in our corpora and in Table \ref{tab:countries} we show  

'ainonline, airwaysmagazine, arstechnica, bleacherreport, breitbartnews, chessbase, cnet, denverpost, foxnews, jaxdotcom, jessicavalenti, jezebel, motherjones, newsobserver, nytimes, phoenixluc, rollcall, seattletimes, sfchronicle, sinow, slate, startelegram, studyfindsorg, thealligator, theathletic, thedailyclimate, thehill, weatherchannel, wired, wsj, yaledailynews'

\begin{table}[t]
    \centering
    \begin{tabular}{lr}
    \toprule
    Language & Count \\
    \midrule
    English & 975 \\
    Spanish, Castilian & 44 \\
    Portuguese & 36 \\
    Nepali & 24 \\
    French & 21 \\
    German & 10 \\
    Japanese & 9 \\
    Norwegian & 8 \\
    Hindi & 7 \\
    Hebrew & 7 \\
    Russian & 7 \\
    Italian & 5 \\
    Ukrainian & 5 \\
    Chinese & 3 \\
    Afrikaans & 3 \\
    Zulu & 2 \\
    Xhosa & 1 \\
    \bottomrule
    \end{tabular}
    \caption{Our corpus comprises homepages from 18 different languages. We assign each news outlet to the language of the majority of it's articles' languages (e.g. the \textit{New York Times} sometimes publishes Spanish-language articles, but is predominantly and English-language newspaper).}
    \label{ref:languages}
\end{table}

\begin{table}[t]
\small
\begin{tabular}{lrr rr}
\toprule
       & \multicolumn{2}{c}{Domain} & \multicolumn{2}{c}{Position x Size} \\ 
\cline{2-5}
Outlet & Train & Test                 & Train & Test \\ 
\midrule
ainonline & 2159 & 540 & 3844 & 962 \\
airwaysmagazine & 1233 & 309 & 1669 & 418 \\
arstechnica & 9349 & 2338 & 17883 & 4471 \\
bleacherreport & 3849 & 963 & 6689 & 1673 \\
breitbartnews & 7824 & 1957 & 15199 & 3800 \\
chessbase & 2094 & 524 & 3151 & 788 \\
cnet & 3769 & 943 & 6521 & 1631 \\
denverpost & 18802 & 4701 & 36607 & 9152 \\
foxnews & 62170 & 15543 & 125096 & 31274 \\
jaxdotcom & 4100 & 1026 & 7206 & 1802 \\
jessicavalenti & 455 & 114 & 512 & 129 \\
jezebel & 6270 & 1568 & 10956 & 2740 \\
motherjones & 2443 & 611 & 3572 & 893 \\
newsobserver & 7538 & 1885 & 14078 & 3520 \\
nytimes & 38432 & 9608 & 74857 & 18715 \\
phoenixluc & 209 & 53 & 305 & 77 \\
rollcall & 6572 & 1643 & 12436 & 3109 \\
seattletimes & 20942 & 5236 & 40882 & 10221 \\
sfchronicle & 5600 & 1401 & 10952 & 2739 \\
sinow & 176 & 44 & 315 & 79 \\
slate & 23527 & 5882 & 45470 & 11368 \\
startelegram & 11964 & 2992 & 22932 & 5734 \\
studyfindsorg & 450 & 113 & 403 & 101 \\
thealligator & 3046 & 762 & 5432 & 1359 \\
theathletic & 22722 & 5681 & 44463 & 11116 \\
thedailyclimate & 7211 & 1803 & 13688 & 3423 \\
thehill & 30800 & 7700 & 59965 & 14992 \\
weatherchannel & 4551 & 1138 & 8094 & 2024 \\
wired & 2058 & 515 & 3196 & 800 \\
wsj & 15569 & 3893 & 30496 & 7625 \\
yaledailynews & 1378 & 345 & 2527 & 632 \\
\bottomrule
\end{tabular}
\caption{Size of training and test sets, in terms of \# of pairs, used in our experiments.}
\end{table}

\begin{table}[t]
\centering
\begin{tabular}{lr}
\toprule
               Country &  Count \\
\midrule
         United States &    892 \\
                Brazil &     37 \\
        United Kingdom &     32 \\
                 Nepal &     24 \\
                Canada &     20 \\
          South Africa &     18 \\
                France &     17 \\
                 Spain &     13 \\
                Mexico &     13 \\
                 India &     10 \\
                 Japan &      9 \\
             Argentina &      9 \\
                Israel &      9 \\
               Germany &      9 \\
                Russia &      8 \\
                Norway &      8 \\
               Ukraine &      6 \\
               Ireland &      6 \\
                 Italy &      5 \\
           New Zealand &      4 \\
               Austria &      3 \\
                Taiwan &      3 \\
              Colombia &      2 \\
             Australia &      2 \\
               Uruguay &      1 \\
                 Qatar &      1 \\
               Belgium &      1 \\
                Latvia &      1 \\
Bosnia and Herzegovina &      1 \\
               Georgia &      1 \\
           El Salvador &      1 \\
               Lebanon &      1 \\
\bottomrule
\end{tabular}
\caption{Countries of origin for the homepages we collect, based on where the organization is based.}
\label{tab:countries}
\end{table}

\section{News Homepage Layouts}
\label{app:news_homepage_layouts}

\begin{figure}[t]
    \centering
    \includegraphics[width=0.6\linewidth]{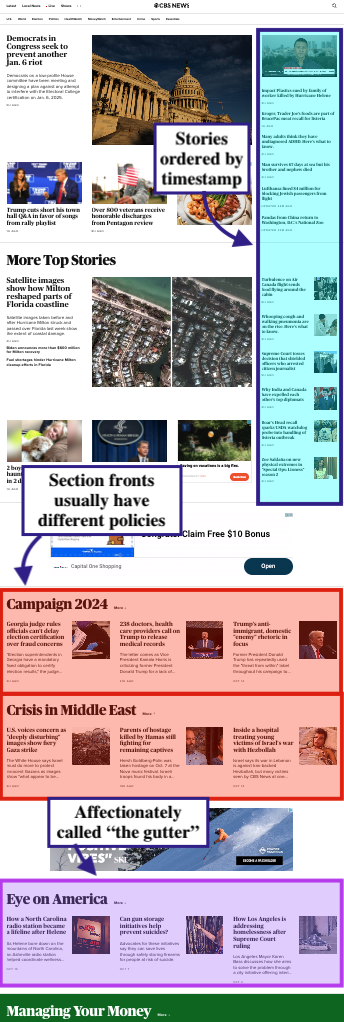}
    \caption{We show three sections of a sample homepage (from CBS News) where editorial decisions for different reasons. We highlight the ``Breaking News'' Section, ``Section Fronts'' and ``The Footer''. }
    \label{fig:annotated-homepage}
\end{figure}

In Figure \ref{fig:annotated-homepage}, we show several areas of a homepage where editorial policies are likely to be unclear and challenging to model. In the top section, a ``Breaking News'' feed shows articles shortly after  they are published. They usually do not stay long in these positions \cite{costanza2016pageonex}, so there is high variability in this section. In the middle section, a ``Section Fronts'' show top articles in each section, each combining the different priorities of the desks \cite{angele2020metrics}. Finally, in this reporter's experience, the bottom of a homepage was affectionately called the ``Gutter''. However, it is more commonly referred to as a ``footer''\footnote{\url{https://alyamanalhayekdesign.com/blog/the-parts-of-a-webpage-a-complete-list/}}

\begin{figure*}[t] 
    \centering
    \includegraphics[width=\linewidth]{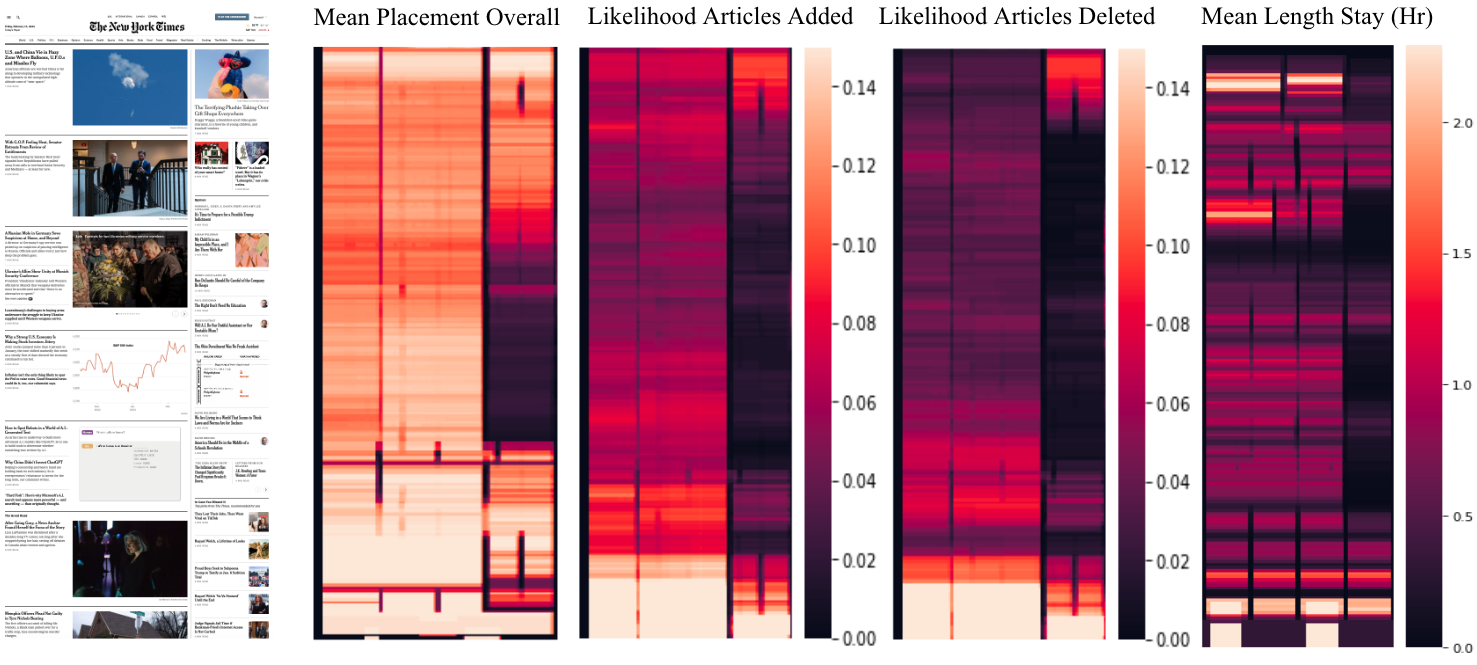}
    \caption{Different analyses we run on bounding boxes across time: average locations of bounding boxes on a homepage, locations where articles are added first, locations where they are removed, and the average time articles in various locations spend.}
    \label{fig:homepage_visualization}
\end{figure*}

\begin{figure*}[t]
    \centering
    \includegraphics[width=.8\linewidth]{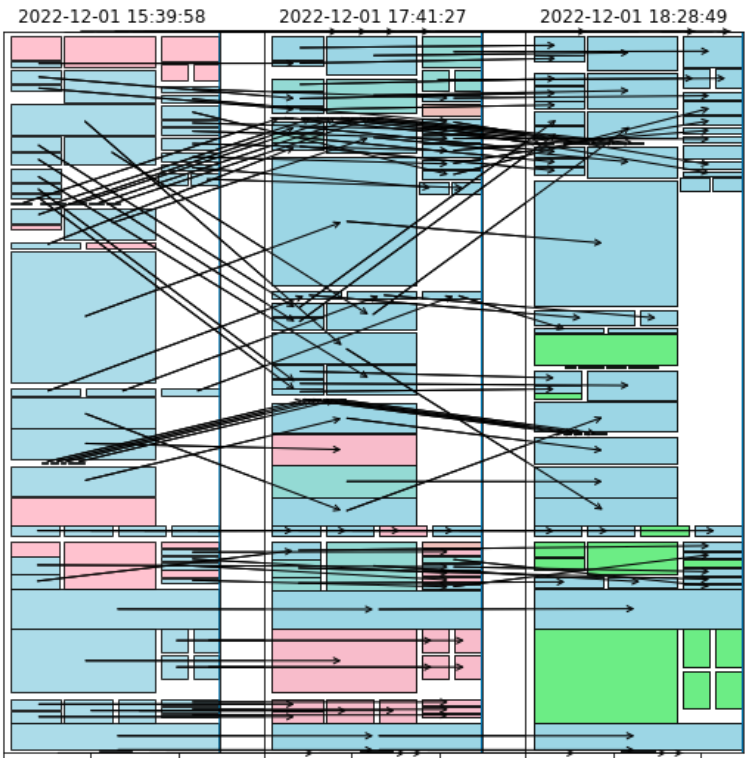}
    \caption{With our suite of tools for parsing homepages, we can examine on a granular level the movement of an article across the homepage.}
    \label{fig:homepage_tracking}
\end{figure*}

We can extend this analysis by visualizing the flow of articles on a homepage over time. We show in Figure \ref{fig:homepage_visualization} where articles tend to get added and deleted overall, as well as where they stay the longest. In Figure \ref{fig:homepage_tracking}, we show how articles shift frequently around a homepage.

\begin{figure}[t]
    \centering
    \begin{subfigure}[t]{0.8\linewidth}
        \centering
        \includegraphics[width=\linewidth]{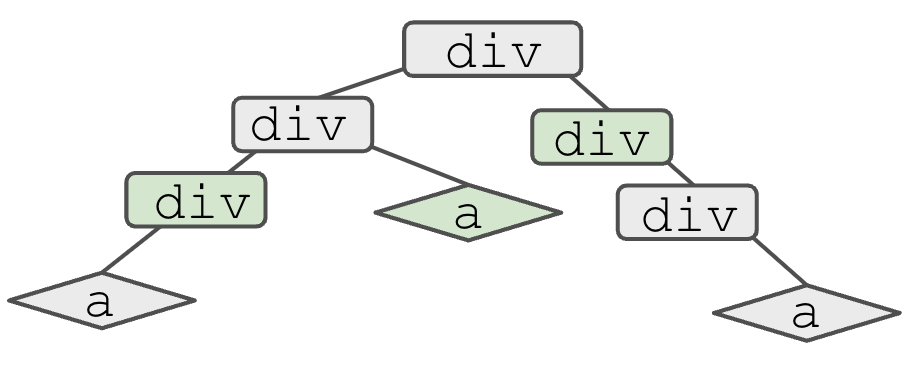}
        \caption{Our deterministic algorithm starts at all $\left<a\right>$ nodes and recursively traverses up the DOM to find maximal subtrees with one $\left<a\right>$. Green nodes shown are article bounding boxes.}
        \label{fig:deterministic_algorithm}
    \end{subfigure}

    \vspace{1em} 
    
    \begin{subfigure}[t]{\linewidth}
        \centering
        \includegraphics[width=.8\linewidth]{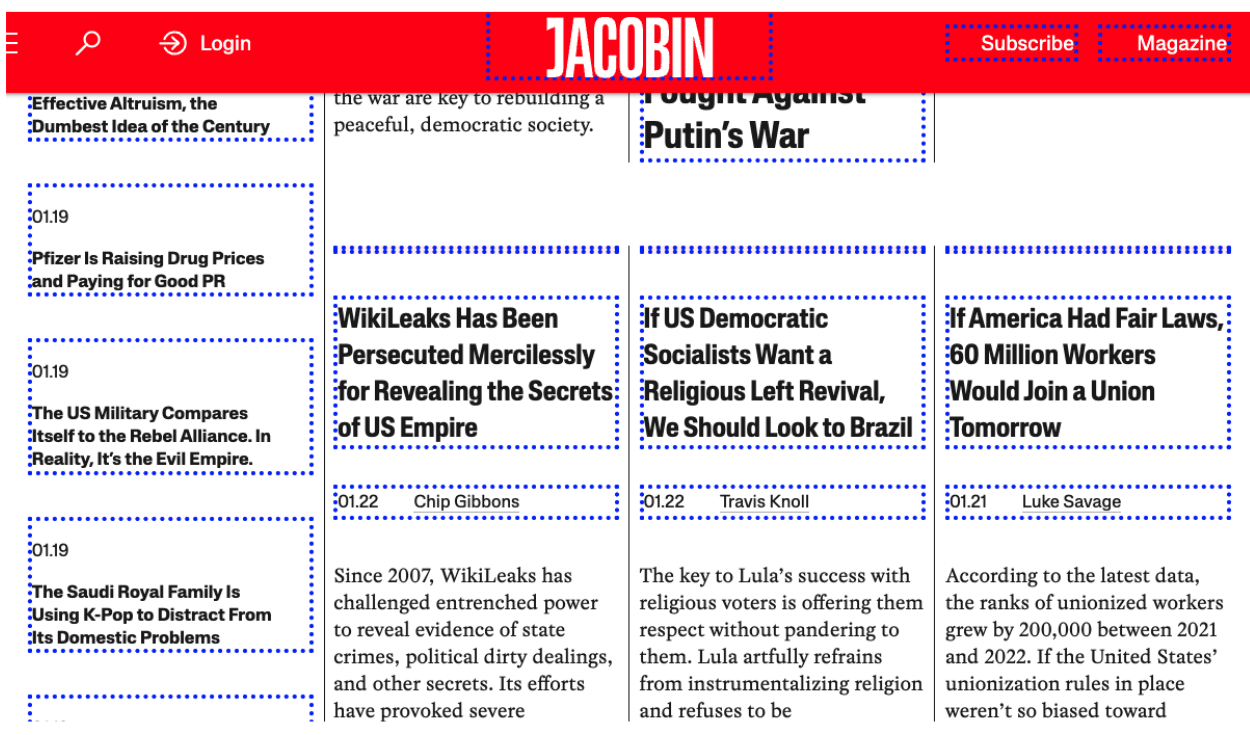}
        \caption{Failure cases (missing text area) with the deterministic algorithm.}
        \label{fig:bounding_box_extraction}
    \end{subfigure}
    \caption{Illustration of our deterministic bootstrapping algorithm and a failure case. Here, when non-article links exist, we misunderstand the full area of an article, excluding the text below.}
    \label{fig:combined_algorithm_figures}
\end{figure}

\begin{figure*}[t] 
    \centering
    \includegraphics[width=\linewidth]{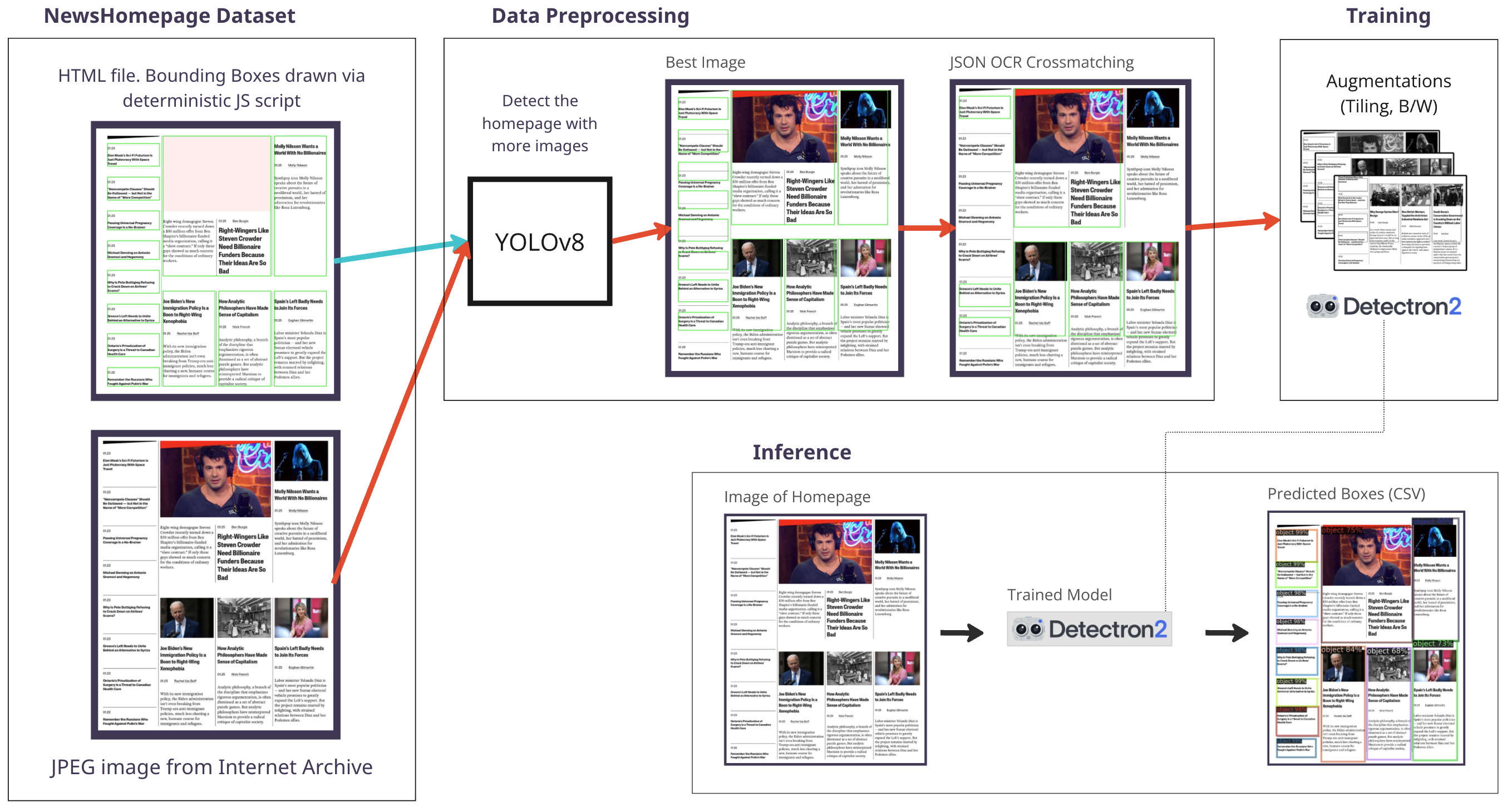}
    \caption{This diagram is an overview of the data preparation and training of the Detectron2 model for predicting bounding box on websites. }
    \label{fig:training-pipeline}
\end{figure*}

\begin{figure*}[t] 
    \centering
    \includegraphics[width=\linewidth]{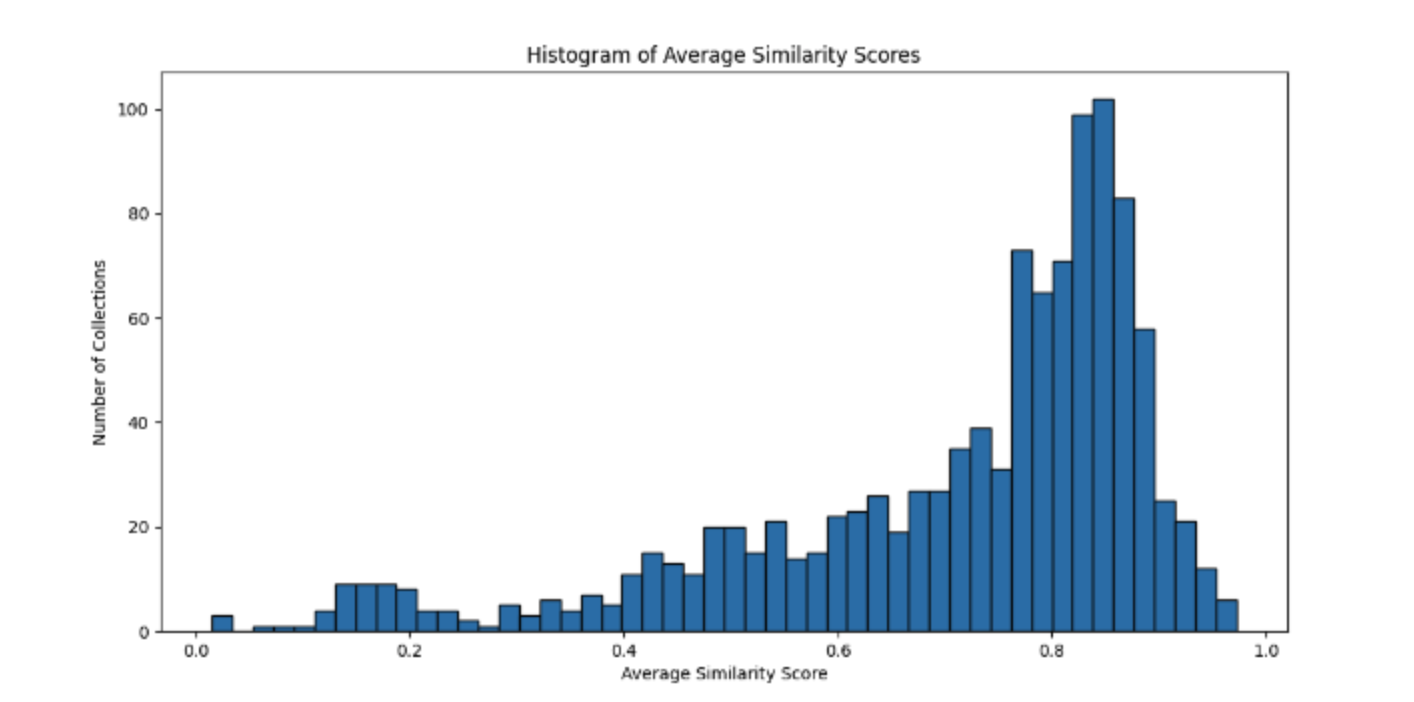}
    \caption{When sorting our sources to determine the ones most difficult for the DOM-Tree algorithm, we define the Average Similarity score to be a general measure as to how well the bounding box's text match the article's JSON file containing text/link pairs. High similarity score means high bounding box accuracy, and vice versa.}
    \label{fig:similarity_histogram}
\end{figure*}
\end{document}